\documentstyle[11pt,epsf]{article}
\renewcommand \baselinestretch{1.3}

\newcommand{\half}{{\scriptstyle{\frac{1}{2}}}}
\newcommand{\BE}{\begin{equation}}
\newcommand{\EE}{\end{equation}}
\newcommand{\msbar}{\overline{\rm MS}}
\begin{document}
\begin{titlepage}
\begin{flushright}
DOE/ER/40717-45 \\
May, 1997
\end{flushright}

\vspace*{14mm}
\begin{center}
{\Large\bf The Banks-Zaks Expansion and \\
\vspace*{4mm}
``Freezing'' in Perturbative QCD}

\vspace*{18mm}
{\Large S.~A. Caveny and P.~M. Stevenson}
\vspace*{5mm}\\
{\large T.~W.~Bonner Laboratory, Physics Department \\
Rice University, Houston, TX 77251, USA}
\vspace{16mm}\\
{\bf Abstract:}
\end{center}

The recent calculation of the four-loop $\beta$-function in QCD 
provides further evidence that the Banks-Zaks expansion in 
$16 \half - n_{f}$ is sufficiently well behaved to be useful even for 
$n_f \approx 2$ light flavours.  This expansion inherently predicts 
``freezing'' of the QCD couplant at low energies as a perturbative effect.  
We consider the $e^+e^-$ and Bjorken-sum-rule cases as examples.
  
\end{titlepage}
 
\setcounter{page}{1}

\section{Introduction}

    The old idea that the QCD running coupling ``freezes'' at low energies 
has been phenomenologically successful in a wide variety of contexts.  
(See Refs. \cite{mat,webber} and references therein.)  Theoretical 
evidence that ``freezing'' does occur, and for purely perturbative reasons, 
comes from the third-order calculation of $R_{e^+e^-}$ \cite{r2} which, 
when ``optimized'' with respect to renormalization scheme, yields such 
behaviour \cite{mat}.  Pad\'e approximant methods indicate a similar 
conclusion \cite{eks}.  Another approach is the Banks-Zaks (BZ) expansion 
\cite{bz,white,grun,st}, within which ``freezing'' is natural and 
ubiquitous.  The relevance of the BZ expansion to low-energy QCD 
phenomenology hinges on an extrapolation in the number of (massless) flavours, 
$n_f$, from $16\half$ down to about 2.  Our point in this paper is that 
the credibility of this extrapolation has been significantly enhanced 
by the recent calculation of the QCD $\beta$ function to 4 loops \cite{rit}.

     The BZ expansion \cite{bz}-\cite{st} is an expansion in 
$n_f^{\rm crit} - n_f$, where $n_f^{\rm crit}$ is the critical number 
of flavours at which asymptotic freedom is lost; {\it i.e.}, where the first 
$\beta$-function coefficient changes sign.  In QCD $n_f^{\rm crit}$ is 
$16 \half$.  For $n_f$ slightly less than $16 \half$ the $\beta$ 
function starts out negative, but quickly turns positive, and there is 
thus an infrared fixed point $a^*$ close to the origin \cite{cas,bz}.  
Provided the couplant $a(\mu)$ initially lies between 0 and $a^*$ 
(necessary for asymptotic freedom), the couplant will ``freeze'' to 
$a^*$ as the energy scale $\mu$ tends to zero.  Thus, in the BZ 
expansion, ``freezing'' is natural and universal and will occur for all 
$n_f < 16 \half$, unless or until the BZ expansion breaks down.  

     Ref. \cite{st} has suggested that the BZ expansion is qualitatively 
relevant to the real world where only 2 or 3 quark flavours are light 
compared to the QCD $\Lambda$ scale.  The crucial issue is whether the 
expansion shows reasonable numerical convergence when $n_f \approx 2$.  
Assuming good behaviour, Ref \cite{st} made a prediction for a certain 
coefficient.  As we discuss, this prediction is borne out by the recent 
calculation of the QCD $\beta$ function to four-loops \cite{rit}.  

     The plan of the paper is as follows:  Section 2 presents the notation 
and extracts the relevant coefficients.  Sections 3 and 4 discuss the 
BZ expansion for quantities at $Q = 0$ and at general $Q$, respectively.  
An important role is played by $\gamma^{*}$, the slope of the $\beta$ 
function at the fixed point, and we discuss its coefficients (the 
``universal invariants'' of Grunberg \cite{grun}) to fourth order.  
Sect. 5 contains concluding remarks.  Appendix A summarizes the 
results for a general $SU(N)$ colour group.  Appendix B discusses 
the issue of renormalization-scheme (RS) invariance and the notion of 
`regular' and `irregular' schemes.  The Higgs-decay case, which we argue 
is not a good guide to infrared behaviour, is discussed in Appendix C.  

\section{Notation}
     We write the $\beta$ function in the form:
\BE
\label{eq: beta}
\beta \left( a\right) \equiv \mu \frac{da}{d\mu} = -b a^2 \left( 1 + ca + 
c_{2} a^2 + c_{3} a^3 +\ldots \right) \,,
\EE
where  $a \equiv  \alpha_{s}/\pi$.  The coefficients, in the $\msbar$ 
scheme, are \cite{cas,c2,rit}:
\begin{eqnarray}
\beta_0 = 2b & \! \! = \! \!& 11 - \frac{2}{3} n_f \,,
\nonumber \\
\beta_1 = 8 bc & \! \! = \! \! & 102 - \frac{38}{3} n_f  \,,
\nonumber \\
\beta_2 = 32 b c_2 & \! \! = \! \! & \frac{2857}{2} - 
\frac{5033}{18} n_f + \frac{325}{54} n_f^2 \,,
\\
\beta_3 = 128 b c_3 & \! = \! & 
\left( \frac{149753}{6} + 3564 \zeta_3 \right) 
- \left( \frac{1078361}{162} + \frac{6508}{27} \zeta_3 \right) n_f 
\nonumber \\
& \!\!\!\!& \mbox{}
+ \left( \frac{50065}{162} + \frac{6472}{81} \zeta_3 \right) n_f^2 
+ \frac{1093}{729} n_f^3.
\nonumber 
\end{eqnarray}
Here $\zeta_s$ is the Riemann zeta-function ($\zeta_3 =1.202056903 \ldots$, 
$\zeta_5 = 1.036927755 \ldots$).  

    For $n_{f}$ just below $16 \half$, the $\beta$ function has a 
zero at $a^{*} \sim -\frac{1}{c}$, and $a^*$ is asymptotically 
proportional to $(16 \half - n_{f})$.  Its limiting form:
\BE
a_{0} \equiv \frac{8}{321}\left( 16 \half - n_{f}\right)
\EE
serves as the expansion parameter for the BZ expansion \cite{st}.  Because 
the constant of proportionality is so small, $a_0$ remains small ($\le 0.4$) 
even with $n_f=0$.  To proceed, one re-writes all perturbative coefficients, 
eliminating $n_f$ in favour of $a_0$.  The first two $\beta$-function 
coefficients, which are RS invariant, become:
\begin{eqnarray}
b &=& \frac{107}{8} a_{0},  \\
c &=& -\frac{1}{a_{0}} + \frac{19}{4}. 
\end{eqnarray}
Within the class of so-called `regular' schemes \cite{grun,st}, which 
includes $\msbar$, perturbative coefficients have a polynomial dependence 
on $n_f$, and we may write 
\BE
\label{ciexp}
c_{i} = \frac{1}{a_{0}} \left( c_{i,-1} + c_{i,0} a_{0} + c_{i,1} a_{0}^2 
+ \ldots \right).
\EE
The coefficients, in $\msbar$, are collected in the table below.

\begin{center}
\begin{tabular}{|rclcr|}
\hline
$c_{1,0}$ &=& $\frac{19}{4}$ &=& 4.75 \\ 
\hline 
$c_{2,-1}$ &$=$& $- \left( \frac{8}{107} \right) 
\left( \frac{37117}{768} \right) $ &=& $-3.61$  \\  
$c_{2,0}$ &=& $\frac{243}{32}$ &=& 7.59 \\ 
$c_{2,1}$ &=& $\left( \frac{107}{8} \right) 
\left(  \frac{325}{192} \right) $ &=& 22.6 \\
\hline
$c_{3,-1}$  &=& $ \left( \frac{8}{107} \right) 
\left( \frac{53981}{1152} + \frac{5335}{32}\zeta_3 \right) $ 
&=& 18.5 \\
$c_{3,0}$ &=& $-\frac{1544327}{13824} - \frac{16171}{288}\zeta_3$ 
&=& $-179$ \\
$c_{3,1}$  &=& $\left( \frac{107}{8} \right) 
\left( \frac{2587}{96} +\frac{809}{144}\zeta_3 \right) $ 
&=& 451 \\
$c_{3,2}$ &=& $ - \left( \frac{107}{8} \right)^2 
\left( \frac{1093}{3456} \right) $ &=& $-56.6$ \\
\hline
\end{tabular}
\end{center}

    The BZ expansion can be applied to any perturbatively calculable
physical quantity of the form:
\BE
\label{eq: R}
{\cal{R}}  = a \left( 1 + r_{1} a + r_{2} a^2 + r_{3} a^3 
+ \ldots\right).  
\EE
In a `regular' scheme the coefficients $r_{i}$ are polynomials in $n_f$, 
and hence in $a_{0}$:
\BE 
\label{riexp}
r_i = r_{i,0} + r_{i,1} a_0 + r_{i,2} a_0^2 + \ldots .
\EE
Note that a term $r_{i,j} a_0^p$ or $c_{i,j} a_0^p$ can be assigned a 
degree $i+j-p$, and all terms in any formula must have matching degree.  

    The prototypical example is the $e^+e^-$ ratio:
\BE
R_{e^{+} e^{-}}(Q) \equiv \frac{\sigma_{\rm tot} \left( e^{+} e^{-} \rightarrow 
{\mbox{\rm hadrons}}\right)}{\sigma \left( e^{+} e^{-} \rightarrow \mu^{+} 
\mu^{-}\right)},
\EE
where, neglecting quark masses, we can write 
$R_{e^{+} e^{-}} \left( Q \right) = 3 
\Sigma q_{i}^2 \left( 1 + {\cal R}_{e^+e^-}\right)$, where ${\cal R}_{e^+e^-}$ 
has the form (\ref{eq: R}).  [Actually, for ${\cal R}_{e^+e^-}$ there is a 
problem in that $r_2$ involves a term 
$1.2395 (\Sigma q_{i})^2/(3 \Sigma {q_{i}}^2)$ whose $n_f$ 
dependence is ambiguous because it depends on the electric charges we 
assign to the additional, ficticious quarks.  This arises because 
$R_{e^+e^-}$ involves not just QCD, but its coupling to electromagnetism.  
Fortunately, this term seems to make only a small numerical contribution.  
We shall ignore it henceforth.]  The coefficients, 
in $\msbar$ with the renormalization scale $\mu$ equated with $Q$, are 
collected in the table below \cite{r2}:

\begin{center}
\begin{tabular}{|rclcr|}
\hline
  & & Coefficients in ${\cal R}_{e^+e^-}$.  [$\msbar$($\mu=Q$)] & & \\
\hline
$r_{1,0}$ &=& $\frac{1}{12}$ &=& 0.0833 \\
$r_{1,1}$ &=& $\left( \frac{107}{8} \right) 
\left( \frac{11}{4} - 2 \zeta_3 \right) $ &=& 4.63 \\  
\hline
$r_{2,0}$ &=& $-\frac{12521}{288} + 13 \zeta_3$ &=& $-27.85$ \\ 
$r_{2,1}$ &=& $ \left( \frac{107}{8} \right) 
\left( \frac{401}{24} - \frac{53}{3}\zeta_3 + \frac{25}{3}\zeta_5 \right) $ 
&=& 55.0 \\
$r_{2,2}$ &=& $ \left( \frac{107}{8} \right)^2 
\left( \frac{151}{18} - \frac{19}{3}\zeta_3 - \frac{\pi^2}{12} \right) $ 
&=& $-8.34$ \\
\hline
\end{tabular}
\end{center}

     Another example is the Bjorken sum rule:
\BE
\int_0^1 dx g_1^{\rm ep-en}(x,Q^2) = \frac{1}{3} 
\left| \frac{g_A}{g_V} \right| \left( 1 - {\cal R}_{\rm Bj} \right).
\EE
(The same QCD corrections, apart from a $(\Sigma q_{i})^2/(3 \Sigma {q_{i}}^2)$ 
term, appear in the Gross Llewellyn-Smith sum rule.)  The coefficients, 
from Ref. \cite{bj} are listed below.

\begin{center}
\begin{tabular}{|rclcr|}
\hline
  & & Coefficients in ${\cal R}_{\rm Bj}$ [$\msbar$($\mu=Q$)] & & \\
\hline
$r_{1,0}$ &=& $- \frac{11}{12}$ &=& $-0.917$ \\
$r_{1,1}$ &=& $\frac{107}{8}$ &=& 13.38  \\  
\hline
$r_{2,0}$ &=& $-\frac{1385}{72} - \frac{55}{4} \zeta_3$ &=& $-35.76$ \\ 
$r_{2,1}$ &=& $ \left( \frac{107}{8} \right) 
\left( \frac{2749}{432} + \frac{61}{18} \zeta_3 - 5 \zeta_5 \right)$ 
&=& 70.25 \\
$r_{2,2}$ &=& $ \left( \frac{107}{8} \right)^2 
\left( \frac{115}{72} \right) $ &=& 285.73 \\
\hline
\end{tabular}
\end{center}

     We mention that the same decomposition of coefficients is needed 
in the ``large-$b$'' approximation \cite{max}, which employs the 
opposite limit ($b \to \infty$, rather than $b = (107/8)a_0 \to 0$ as here).  

\section{BZ Expansion: Q = 0}

     The fixed-point condition $\beta(a^{*}) = 0$ always has a solution 
as a power series in $a_0$:
\BE
\label{astar}
a^{*} = a_{0} \left[ 1 + v_1 a_{0} + v_2 a_{0}^{2} + 
v_3 a_{0}^{3} + \ldots \right].
\EE
A straightforward calculation yields:
\begin{eqnarray}
\label{eq: AA}
v_1 &=& c_{1,0} +c_{2,-1},\nonumber \\
v_2 &=& (c_{1,0}+2c_{2,-1})(c_{1,0} +c_{2,-1})+c_{2,0}+c_{3,-1}, \\  
v_3 &=&
c_{1,0}^3 + 6 c_{1,0}^2 c_{2,-1} + 
c_{1,0}(3 c_{2,0} + 4 c_{3,-1} + 10 c_{2,-1}^2)
\nonumber \\
& & {\mbox{}} + c_{2,-1}(4 c_{2,0} + 5 c_{3,-1}) + 5 c_{2,-1}^3 + c_{2,1} +
c_{3,0} + c_{4,-1} \,. \nonumber
\end{eqnarray}
(Numerically, $v_1 =1.1366$, $v_2 =23.27$, $v_3 = c_{4,-1} - 138.6$, 
in the $\msbar$ scheme.  The poor apparent convergence of the $a^*$ 
series need not bother us, since $a^*$ is RS dependent.)

     A physical quantity ${\cal R}$ will also have an infrared 
limit given by a power series in $a_{0}$.  One simply takes the 
perturbative expansion of ${\cal R}$, Eq. (\ref{eq: R}); substitutes 
$a=a^*$, given by (\ref{eq: AA}); and re-expands in powers of $a_{0}$.  
This yields:
\BE
\label{eq: RR}
{\cal R}^{*} = a_{0} \left[ 1 + w_1 a_{0} + w_2 a_{0}^{2} + 
w_3 a_{0}^{3} + {\cal O}(a_{0}^{4}) \right],
\EE
where
\begin{eqnarray}
w_1 &=& v_1 +r_{1,0},\nonumber \\
w_2 &=& v_2 + 2r_{1,0} v_1 + r_{2,0} + r_{1,1}, \\
w_3 &=& v_3 + (2 v_2 + v_1^2) r_{1,0} + v_1 (2r_{1,1} + 3 r_{2,0})
+ r_{2,1}+ r_{3,0}.\nonumber 
\end{eqnarray}
These coefficients are RS-scheme independent (see Appendix B) and so their 
numerical values are significant.  They should be order-1 numbers, if 
all is to be well.  

      For the $e^+e^-$ case, Ref. \cite{st} obtained the value of the 
first coefficient, $w_1 = 1.22$, but $w_2$ could only be obtained as 
$-18.25 + c_{3,-1}$, since $c_{3,1}$ was then unknown.  To quote 
Ref. \cite{st}: ``For the expansion to be credible one needs 
$c_{3,-1}(\msbar)$ to be in the range, say, $+13$ to $+21$.''  This 
prediction is confirmed by the new $\beta$-function result \cite{rit}, 
which yields $c_{3,-1} = 18.5$.  Therefore, $w_2$ is quite small, 
$0.23$, giving a respectable series:
\BE 
\label{ree}
{\cal R}_{e^+e^-}^* = a_0 \left[ 1 + 1.22 a_0 + 0.23 a_0^2 + \ldots \right].
\EE
The next coefficient, 
\BE 
\label{wee}
w_3{\scriptstyle{(e^+e^-)}} 
= c_{4,-1} + r_{3,0}{\scriptstyle{(e^+e^-)}} - 164.0,
\EE
would require calculation of both $\beta$ and ${\cal R}_{e^+e^-}$ to one 
more order.

      In the Bj-sum-rule case the corresponding result is
\BE
\label{rbj}
{\cal R}_{\rm Bj}^* = a_0 \left[ 1 + 0.22 a_0 - 1.21 a_0^2 + \ldots \right] ,
\EE
where the coefficients are also of order unity.  It is interesting that 
the ${\cal R}^*$ in this case is even smaller than in the $e^+e^-$ case.  
The next coefficient is 
\BE 
\label{wbj}
w_3{\scriptstyle{({\rm Bj})}} 
= c_{4,-1} + r_{3,0}{\scriptstyle{({\rm Bj})}}  - 203.7.
\EE

\section{BZ Expansion: Nonzero Q}

    A formulation of the BZ expansion for quantities at a general $Q$ 
was derived in Ref. \cite{st}.  We briefly review the main ingredients.  
First, we need a suitable form of the boundary condition for the 
$\beta$-function equation.  Setting 
$\hat{\beta} \left( x \right) \equiv \beta \left( x \right) / b$, we 
write
\BE
\label{eq: int} 
b \ln \left( \frac{\mu}{\tilde{\Lambda}} \right) = \lim_{ \delta \rightarrow 
0} \left[ \int_{\delta}^{a} \frac{dx}{\hat{\beta} \left(x\right)} + 
{\cal C}\left( \delta\right)\right].  
\EE
The constant of integration \({\cal C}\left( \delta\right)\) needs to be 
suitably singular as $\delta \rightarrow 0$ and we choose \cite{opt,st}:
\begin{eqnarray}
{\cal C} \left( \delta \right) &=& \mbox{\rm P.V.} \int_{\delta}^{\infty} 
\frac{dx}{x^2\left(1+cx\right)} \nonumber \\
&=& \frac{1}{\delta} + c\ln\delta + c \ln \left|c\right| + 
{\cal O}\left(\delta\right).
\end{eqnarray}
Note that Cauchy's principal value (P.V.) is introduced to deal with the 
pole at $x = -1/c$  when $c < 0$.  This choice amounts to a definition of 
$\tilde{\Lambda}$, within a given RS.  [We use a tilde to distinguish it 
from the older, but still widely used, definition of the $\Lambda$ 
parameter \cite{bbdm}.  The relation is $\ln(\Lambda/\tilde{\Lambda}) = 
(c/b) \ln(2 \! \mid \! c \! \mid \! /b)$.  While the two definitions are not 
dissimilar for small $n_f$, they become infinitely different as 
$n_f \to 16 \half$.  In the BZ-expansion context the use of $\tilde{\Lambda}$ 
is much more convenient.]

    As explained in Ref. \cite{st}, it is convenient to put the $\beta$ 
function into the form 
\BE
\label{betaf}
\frac{1}{\hat{\beta}\left(x\right)} = - \frac{1}{x^2} + \frac{c}{x} - 
\frac{1}{ {\hat{\gamma}}^{*} \left(a^{*} - x\right)} +  H\left(x\right).
\EE
where ${\hat{\gamma}}^{*}$ is $\gamma^{*}/b$, with $\gamma^{*}$ being 
the slope of the $\beta$ function at the fixed point:
\BE
\label{gamstar}
\gamma^{*} \equiv \left. \frac{\beta\left(x\right)}{dx}\right|_{x=a^{*}} 
= -b a^{*} 
\left( 1 + 2 ca^{*} + 3 c_{2} {a^{*}}^{2}+ 4 c_{3} {a^{*}}^{3} +\ldots \right).
\EE
As discussed below, $\hat{\gamma}^*$ can be obtained as a series in $a_0$.  
The remainder function $H(x)$ can be expanded as a power series, 
$H_0 + H_1 x + \ldots$, whose coefficients are of order $a_0$.  

    One now inserts (\ref{betaf}) into (\ref{eq: int}) and performs the 
integration.  One can then eliminate $a$ and $a^{*}$ in favour of 
${\cal R}$ and ${\cal R}^{*}$.  In fact, since the result must be RS 
invariant, one can --- without loss of generality --- short-cut this 
step by utilizing the ``effective-charge'' RS in which $a \equiv {\cal R}$.  
This leads to the formula 
\cite{st}:
\BE 
\label{eq: YY}
\rho_{1} = \frac{1}{\cal R} + \frac{1}{\hat{\gamma}^{*(n)}} 
\ln\left(1- \frac{{\cal R}}{{\cal R}^{*(n)}} \right) 
+ c\ln\left(\left|c\right| {\cal R} \right) 
+ \sum_{i=0}^{n-4} \frac{H_i^{\rm (ec)} {\cal R}^{i+1}}{i+1}.
\EE
The last term, involving the $H_i^{\rm (ec)}$ coefficients (of the 
effective-charge scheme), is only relevant in fourth order and beyond.  
Thus, for the first three orders the equation takes the same form, just with 
the parameters $\hat{\gamma}^*$ and ${\cal R}^*$ approximated to the 
appropriate order.  On the left-hand side, $\rho_1$ is the RS invariant 
\cite{opt}
\BE
\label{eq: X}
\rho_{1} \equiv b\ln\left( \frac{\mu}{\tilde{\Lambda}}\right) - r_{1} 
\equiv b \ln\left( \frac{Q}{\tilde{\Lambda}_{\rm eff}} \right),
\EE
where $\tilde{\Lambda}_{\rm eff}$ is a characteristic scale specific to 
the particular physical quantity ${\cal R}$.  It is related to the 
$\tilde{\Lambda}$ parameter of some reference scheme (eg. $\msbar$) by 
an exactly calculable factor $\exp(r_1/b)$ involving the $r_1$ coefficient 
in that scheme, evaluated at $\mu=Q$.  (We caution that $\rho_1$ {\it cannot} 
be split into ${\cal O}(1)$ and ${\cal O}(a_0)$ pieces in a RS-invariant 
way \cite{st}.)

    Numerically inverting Eq. (\ref{eq: YY}) provides ${\cal R}$ as a 
function of $Q$.  The resulting ${\cal R}(Q)$ naturally agrees with ordinary 
perturbation theory to the corresponding order at large $Q$, but freezes 
to the value ${\cal R}^{*(n)}$ as $Q \to 0$.  The BZ series expansion for 
${\cal R}^{*}$ was discussed in the previous section.  The BZ expansion 
for ${\hat{\gamma}}^*$ is obtained straightforwardly by substituting the 
expansion of $a^*$ (Eqs. (\ref{astar}) and (\ref{eq: AA})) into 
(\ref{gamstar}).  This gives:
\BE
{\hat{\gamma}}^{*} = a_{0}\left[ 1 + g_1 a_{0} + 
g_2 {a_{0}}^{2} + g_3 {a_{0}}^{3} + \ldots \right]
\EE
where,
\begin{eqnarray}
g_1 &=&  c_{1,0},\nonumber 
\\
g_2 &=& {c_{1,0}}^{2} - {c_{2,-1}}^{2} -c_{3,-1} 
, \\
g_3 &=&{c_{1,0}}^{3} - {4c_{2,-1}}^3 
- 5 {c_{1,0}}{c_{2,-1}}^{2} - 4 {c_{1,0}} c_{3,-1}
\nonumber \\
& & \mbox{} - 2 c_{2,-1} c_{2,0} - 6 c_{2,-1} c_{3,-1} - c_{3,0} - 2c_{4,-1}.
\nonumber
\end{eqnarray}
It is noteworthy that certain terms of degree $n$ are absent in $g_n$:  
$g_1$ does not contain $c_{2,-1}$; $g_2$ does not contain $c_{2,0}$ or 
$c_{2,-1} c_{1,0}$; and $g_3$ does not contain $c_{2,1}$ or $c_{2,0} c_{1,0}$ 
or $c_{2,-1} {c_{1,0}}^2$.  

      The significance of $\gamma^* = b \hat{\gamma}^*$ is that it 
is the `critical exponent' governing how ${\cal R}$ approaches 
${\cal R}^{*}$ as $Q \rightarrow 0$; asymptotically, 
$ {\cal R} - {\cal R}^* \propto Q^{\gamma^{*}}$.  
As pointed out by Grunberg, the $g_n$ coefficients are RS invariants, 
and are universal, in the sense that they are not specific to some 
particular physical quantity ${\cal R}$.  (See Appendix B for discussion 
of some subtleties.)

      The new $\beta$-function result \cite{rit} enables us to determine 
the numerical value of the second invariant; $g_2 =-8.99$.  (The exact 
expression, for general $N$, is given in Appendix A.)  Hence the 
${\hat{\gamma}}^*$ series is:
\BE
\label{gamser}
{\hat{\gamma}}^{*} = a_{0}\left[ 1 + 4.75 a_{0} - 8.99 {a_{0}}^{2} 
+ \ldots \right]
\EE
Clearly, the $\hat{\gamma}^*$ series is not as well behaved as the 
${\cal R}^*$ series that we saw earlier.  In Fig. 1(a) we show $\gamma^*$ 
($= b \hat{\gamma}^*$) as a function of $n_f$.  (Fig. 1(b) shows 
shows the same quantity normalized by $1/(ba_0)$.)  The lower and upper solid 
curves are the first- and second-order results, respectively, while the 
middle solid curve is the third-order result.  The dashed curve represents 
the third-order result re-cast as a Pad\'e approximant:
\BE 
\label{gampade}
\hat{\gamma}^* \approx  a_0 \frac{(1 + 6.64 a_0)}{(1 + 1.89 a_0)}.
\EE
For comparison we also give, as the dotted curve, the prediction arising 
from an optimized-perturbation-theory analysis of the $e^+e^-$ case 
\cite{mat} (see comments in Appendix B).  The reasonable agreement between 
the last three curves gives us some confidence that the extrapolation 
to low $n_f$ is qualitatively valid, even if the quantitative precision is 
not good.  

      The next coefficient is 
\BE
\label{g3}
g_3 = 269.44 - 2c_{4,-1}.
\EE
We therefore predict that $c_{4,-1}(\msbar)$ will turn out to 
be somewhere around $135 \pm 10$.  Assuming that the $w_3$ coefficients 
in (\ref{wee}) and (\ref{wbj}) are modest, we can expect that 
$r_{3,0}{\scriptstyle{(e^+e^-)}} \approx 29 \pm 10$ and 
$r_{3,0}{\scriptstyle{({\rm Bj})}} \approx 68 \pm 10$.  

    Knowing $\hat{\gamma}^*$ and ${\cal R}^*$, we can numerically solve Eq. 
(\ref{eq: YY}) to obtain ${\cal R}(Q)$ as a function of 
$Q/\tilde{\Lambda}_{\rm eff}$.  The result, for the $e^+e^-$ case in 
third order, is shown in Fig. 2 for $n_f = 14, 10, 6$ and $2$.  
(It may be directly compared with similar figures for first and second 
order in Ref. \cite{st}.)  Corresponding results in the Bjorken-sum-rule 
case are shown in Fig. 3.  Note that the $\tilde{\Lambda}_{\rm eff}$'s 
defining the units of $Q$ in Figs. 2 and 3 are not the same: however, 
they are easily converted to a common $\tilde{\Lambda}$ using 
Eq. (\ref{eq: X}).

\section{Concluding Remarks}

The result of van Ritbergen {\it et al.}'s \cite{rit} calculation of the 
$\beta$ function to four-loops sheds much light on the BZ expansion. 
It supports the idea that the expansion is relevant to the phenomenologically 
interesting case of only two light quark flavors.  That, in turn, 
implies perturbatively explicable ``freezing'' of the QCD running coupling 
constant.  

   We mention again the very interesting work on the ``large-$b$'' 
approximation (see \cite{max} and references therein) which draws on 
large-$n_f$ results.  This is the opposite approximation to ours.  
It extrapolates upwards from $n_f = - \infty$ ({\it minus} 
infinity if the theory is to be asymptotically free) towards $n_f \approx 2$, 
whereas we are extrapolating down from $n_f = 16\half$.  We think both 
approximations are useful; neither is very precise, but both seem to be 
qualitatively valid, and offer a great deal of insight into QCD.  Our 
preliminary studies indicate that the large-$b$ approximation also 
predicts ``freezing,'' and we hope to report on this shortly.

\vspace*{4mm}

{\bf Acknowledgements}:  We thank Ji\v{r}\'i Ch\'yla for correspondence, 
particularly regarding the issues in Appendix B.  
This work was supported in part by the U.S. Department of Energy under
Grant No. DE-FG05-92ER40717.

\newpage

\section*{Appendix A: SU($N$) generalization}

    The critical number of flavours is:
\BE
n_f^{\rm crit} = \frac{11}{2} N.
\EE
The $\beta$-function coefficients in $\msbar$ are \cite{rit}
\begin{eqnarray}
\beta_0 = 2b & \! = \!&\frac{1}{3}(11 N - 2 n_f) 
\nonumber \\
\beta_1 = 8 bc & \! = \! & \frac{1}{3 N}(34 N^3 - n_f (13 N^2 - 3))  
\nonumber \\
\beta_2 = 32 b c_2^{\msbar} & \! = \! & \frac{1}{108 N^2} \left( 5714 N^5 + 
n_f(-3418 N^4 + 561 N^2 + 27) + n_f^2(224 N^3 - 66 N) \right)
\nonumber \\
\beta_3 = 128 b c_3^{\msbar} & \! = \! & \frac{1}{1944 N^3} 
\left( 601892 N^7 - 25920 N^5  + \zeta_3 (9504 N^7 + 684288 N^5)
\right.
\nonumber \\
& \!\! & \mbox{} + n_f \left[ - 485513 N^6 + 58583 N^4 - 21069 N^2 - 5589 
\right.
\nonumber \\
& \!\! & \mbox{} + \left. 
\zeta_3 (- 4320 N^6 - 118368 N^4 + 9504 N^2) \right] 
\nonumber \\
& \!\! & \mbox{} + n_f^2  \left[ 69232 N^5 - 19816 N^3 -22428N 
+ \zeta_3 (18144 N^5 - 13824 N^3 + 52704N ) \right]
\nonumber \\
& \!\! & \mbox{} + \left. n_f^3 [1040 N^4 - 616 N^2] \right)
\end{eqnarray}
The BZ expansion parameter becomes 
\BE 
a_0 = \frac{16}{3(25 N^2 -11)} (n_f^{\rm crit} - n_f)
\EE
The invariant coefficients are 
\begin{eqnarray}
b &\!=\!& \frac{(25 N^2-11)}{16} a_{0},  \\
c &=& -\frac{1}{a_{0}} + \frac{(13N^2-3)}{8 N}. 
\end{eqnarray}
In $\msbar$
\begin{eqnarray}
c_{2,-1} &\! =\! & \frac{(-1402 N^4 + 242 N^2 + 33)}{48 N (25 N^2 - 11)}, 
\\
c_{2,0} &\! = \! & \frac{(318N^4 + 55 N^2 -9)}{384 N^2},
\\
c_{2,1} &\! = \! & \frac{(25 N^2 -11) (112 N^2 - 33)}{3072 N}
\\
c_{3,-1} &\! = \! & \frac{(14731 N^6 -30047N^4 - 58839 N^2 - 2277)}
{1152 N^2(25N^2 -11)} 
\nonumber \\
& \! \! & \mbox{} + 
\frac{11}{8}\frac{(25 N^4 - 18 N^2 + 77)}{(25N^2-11)} \zeta_3
\end{eqnarray}

    The Grunberg invariants, the coefficients in the expansion of 
$\hat{\gamma}^*$ are: 
\begin{eqnarray}
g_1 & \! \! = \! \! & \frac{(13N^2-3)}{8 N},
\\
g_2 & \! \! = \! \! &
\frac{(366782 N^8 - 865400 N^6  + 1599316 N^4  - 571516 N^2 -3993)}
{768 N^2  (25 N^2 - 11)^2}
\nonumber \\
& \!\!\!\! & \mbox{} 
- \frac{11}{8} \frac{(25 N^4 - 18 N^2 + 77)}{(25 N^2 - 11)} \zeta_3. 
\end{eqnarray}

\section*{Appendix B: Regular and Irregular Schemes}

     In this appendix we discuss renormalization-scheme invariance of 
the BZ expansion results.  Since the expansion parameter 
$a_{0} = \frac{8}{321} (16\frac{1}{2} - n_{f})$ is an RS-invariant pure 
number, one expects the coefficients in the BZ expansion of a physical 
quantity ${\cal R}$ to be RS-invariant, but it is important to have 
confirmation.  We first consider the fixed point results at $Q=0$ (which 
are independent of the $\tilde{\Lambda}$ parameter).  

    In the text, as in Ref \cite{st}, we limited the discussion to 
so-called `regular schemes' in which the coefficients in ${\cal R}$ 
and $\beta(a)$ have a polynomial dependence on $n_f$.  Such schemes 
are natural in diagrammatic terms, since each fermion loop gives an 
$n_f$ factor.  They are convenient for our purposes, since the 
BZ-expansion coefficients are easily extracted from calculations 
made in those schemes.  However, we emphasize that ``irregular'' schemes 
are not necessarily ``bad.''  They will also lead to the same BZ-expansion 
results, but the extraction of the BZ coefficients from calculations in those 
schemes will be less straightforward.  An analogous situation arises, 
for example, with gauge invariance; a certain class of gauges may 
be convenient for some purposes, but inconvenient for others.  

    It is easy to show the invariance of the BZ coefficients within the 
class of `regular' schemes.  This was done in Ref. \cite{st} by 
considering the RS-invariants $\rho_2, \rho_3, \ldots$ \cite{opt} 
that are invariant combinations of ${\cal R}$ and $\beta(a)$ coefficients, 
and expanding them in powers of $a_0$.  A `low-brow' version of the proof 
is also instructive.  Consider a change of RS, $ a \mapsto a'$, where
\BE
\label{rstrans}
a' = a\left( 1 + u_{1} a + u_{2} a^{2} + \ldots \right).
\EE
The coefficients in the expansion of ${\cal R}$ change to 
\begin{eqnarray}
r'_{1} &=& r_{1} - u_{1}\, , \nonumber \\
r'_{2} &=& r_{2}- u_{2} -2 u_{1}r_{1} +2u_{1}^{2} \, , 
\end{eqnarray}
etc..  The $\beta$ function transforms to 
$\beta'(a') =(\partial a'/\partial a) \beta(a)$, whose coefficients are: 
\begin{eqnarray}
c' &=& c \nonumber \\
c'_2 &=& c_2 + u_2 -u_1^2 - u_1 c, \nonumber \\
c'_3 &=& c_3 + 2 u_3 + c u_1^2 - 2 c_2 u_1 - 6 u_1 u_2 + 4 u_1^3, 
\end{eqnarray}
etc..  If both the primed and unprimed scheme are regular, then the 
$c_i$, $c_i'$ and $r_i$, $r_i'$ coefficients are expandable as in 
Eqs. (\ref{ciexp}), (\ref{riexp}), and the $u_i$ coefficients in the 
scheme transformation can be expanded as:
\begin{eqnarray}
\label{uiexp}
u_{1} &=& u_{1,0} + u_{1,1}a_0 \, , \nonumber \\
u_{2} &=& u_{2,0} + u_{2,1}a_0 + u_{2,2}a_0^{2}\, , 
\end{eqnarray}
etc..  It is then straightforward to prove the invariance of the combinations 
appearing in the BZ expansion of ${\cal R}$ and $\gamma^*$.  For instance, 
from the relations 
\begin{eqnarray}
r'_{1,0} &=& r_{1,0} - u_{1,0} \, , \nonumber \\
c'_{2,-1} &=& c_{2,-1} + u_{1,0} \, ,   \nonumber \\
c'_{3,-1} &=& c_{3,-1} - u_{1,0}^2 - 2 c_{2,-1} u_{1,0}, 
\end{eqnarray}
one sees that $r'_{1,0} + c'_{2,-1} = r_{1,0} + c_{2,-1}$, 
and $c'_{3,-1} + (c'_{2,-1})^2 = c_{3,-1} + (c_{2,-1})^2$, showing that 
these combinations are RS invariant.  Extending this procedure one can prove 
the invariance of the higher-order $w_i$ and $g_i$ coefficients.  

     Any scheme related to $\msbar(\mu=Q)$ by a transformation (\ref{rstrans}) 
with $u_i$'s expandable as in (\ref{uiexp}) is a `regular' scheme.  
[It is noteworthy that in a general, regular scheme the coefficients 
$c_{2,2}, c_{3,3}, \ldots$ are non-zero, while in a more restrictive class 
of `strictly regular' schemes, of which $\msbar(\mu=Q)$ is an example, 
these coefficients vanish.  This distinction is unimportant for the BZ 
expansion, but matters for the large-$b$ approximation \cite{max}.]  The 
`effective charge' (or `FAC') scheme, in which $a \equiv {\cal R}$ for 
some specific physical quantity, is a `regular' (but not `strictly regular') 
scheme.  However, it is easy to construct RS's that are `irregular' simply 
by considering a transformation in which the $u_i$ depend on $n_f$ in a 
non-polynomial fashion, so that the coefficients $r'_i$ are no longer 
expandable in positive powers of $a_0$.  There is nothing intrinsically 
`bad' about such schemes.  They can arise quite naturally.  For example, 
the 't Hooft scheme, in which $c_2=c_3=\ldots=0$, is `irregular' \cite{max}.  
[Recall that, in `regular' schemes, $c_{3,-1} + (c_{2,-1})^2$ is invariant 
and does not vanish.]  The principle-of-minimal-sensitivity (PMS) scheme 
for any given physical quantity is also `irregular.'  In both these 
cases the $r_i$ coefficients have $1/a_0$ pieces.  

    In a `regular' scheme, obtaining the BZ expansion to $n$th order requires 
terms of order $n+1$ in the $\beta$ function, and of order $n$ in the 
physical quantity ${\cal R}$.  In an `irregular' scheme the same information 
is distributed among higher-order coefficients as well.  Starting from an 
`irregular' scheme, one would require some knowledge of perturbative 
coefficients to higher orders; maybe to all orders.  However, provided 
one carefully kept all terms that could contribute to a given order in $a_0$, 
one would obtain the same BZ expansion.  

    Finally, we discuss the finite-$Q$ case.  Following Ref. \cite{st}, 
we have formulated the result as Eq. (\ref{eq: YY}), which is to be solved 
numerically to obtain ${\cal R}$ as a function of $Q$.  This formula 
involves the RS-invariant quantities $\rho_1$,, $\hat{\gamma}^*$, 
${\cal R}^*$, $c$, and $H_i^{\rm (ec)}$.  The $\tilde{\Lambda}$ parameter 
and $Q$ appear only in $\rho_1$.  The $H_i^{\rm (ec)}$ coefficients 
(relevant only in 4th order and beyond) are directly related to the 
$\tilde{\rho}_2, \tilde{\rho}_3, \ldots$ invariants of Ref. \cite{opt}
which can be conveniently redefined (hence the tilde) to coincide with the 
$\beta$-function coefficients of the `effective charge' scheme \cite{exinv}.  

    The RS invariance of the BZ expansion of $\hat{\gamma}^*$ \cite{grun} 
is verifiable by the procedure discussed above.  It is expected because 
of the well-known result \cite{gp} that the slope of the $\beta$ function 
at a fixed point is an invariant.  However, there is an important 
caveat to the last statement \cite{chyla}, which necessitates some 
further discussion.  The quoted result follows by differentiating the 
$\beta$-function transformation,
$\beta'(a') =(\partial a'/\partial a) \beta(a)$, to give
\BE
\frac{\partial \beta'}{\partial a'} = 
\frac{\partial a}{\partial a'} 
\frac{\partial^2 a'}{\partial a^2} \beta(a) 
+ \frac{\partial \beta}{\partial a}.
\EE
The first term vanishes at the fixed point --- provided that neither 
$\partial a/\partial a'$ nor $\partial^2 a'/\partial a^2$ 
is singular there \cite{gp}.  Ch\'{y}la \cite{chyla} has pointed out that, 
in general, it can be quite natural for those factors to be singular (and 
arbitrary scheme transformation can of course make fixed points appear 
and disappear!).  In general, then, the critical exponent $\gamma^*$ in 
${\cal R} - {\cal R}^* \propto Q^{\gamma^{*}}$ as $Q \to 0$ is {\it not} 
the same as $\partial \beta/\partial a|_{a=a^*}$.  However, these two 
quantities will coincide in a large class of schemes.  It also seems 
safe to assume that they coincide in the context of the BZ expansion, 
where there is necessarily a fixed point at $a^* = a_0 + \ldots$.  

    [We find that the 3rd-order PMS scheme, though `irregular,' also 
yields $\gamma^* = \partial \beta/\partial a|_{a=a^*}$.  This requires 
a detailed analysis of the optimization equations \cite{opt,mat} as 
$Q \to 0$.  The PMS results (in the $e^+e^-$ case) for $\gamma^*$ are 
shown as the dotted curve in Fig. 1.]  

\section*{Appendix C: Higgs decay}

     In this appendix we discuss the case of Higgs-boson decay into 
hadrons.  This seems to be a much more problematic than the cases 
discussed earlier.  It could be viewed as conflicting with the 
general picture we have presented.  We shall argue, though, that the 
crucial role of quark masses in this quantity makes it unsuitable as 
a guide to the infrared behaviour of perturbative QCD.  First, let 
us discuss the numbers.  

     The hadronic decay width of the Higgs has the form 
$\Gamma_H = \frac{3G_F}{4\sqrt{2}\pi} M_H \sum_q m_q^2 \Gamma(a)$ 
with $\Gamma(a) = 1 + \Gamma_1 a + \ldots$, and where $m_q$ is the 
running quark mass evaluated at some scale.  For $\Gamma_H$ there 
is a factorization-scheme ambiguity (how much of the radiative 
corrections should be absorbed into $m_q^2$, and much should be 
left in the explicit series $\Gamma(a)$?).  However, one can define 
the quantity \cite{ver}
\BE
{\cal R}_{\rm Higgs} = -\frac{1}{2} \frac{d \ln(\Gamma_H/M_H)}{d \ln M_H^2} 
\EE
which is free of this factorization-scheme ambiguity, and is a 
physical quantity of the same form ${\cal R} = a(1+ r_1 a + \ldots )$ 
considered earlier.  The coefficients, from Ref. \cite{ver} are collected 
in the table below.

\begin{center}
\begin{tabular}{|rclcr|}
\hline
  & & Coefficients in ${\cal R}_{\rm Higgs}$.  [$\msbar$($\mu=Q$)] & & \\
\hline
$r_{1,0}$ &=& $\frac{23}{12}$ &=& $1.917$ \\
$r_{1,1}$ &=& $\left( \frac{107}{8} \right) \left( \frac{11}{6} \right) 
$ &=& $24.52$ \\  
\hline
$r_{2,0}$ &=& $-\frac{503}{18} - \frac{55}{4} \zeta_3$ &=& $-44.47 $ \\ 
$r_{2,1}$ &=& $ \left( \frac{107}{8} \right) 
\left( \frac{2935}{144} - \frac{7}{4}\zeta_3 - \frac{\pi^2}{3} \right) $ 
&=& $200.5$ \\
$r_{2,2}$ &=& $ \left( \frac{107}{8} \right)^2 
\left( \frac{275}{72} - \zeta_3 - \frac{\pi^2}{12} \right) $ 
&=& $321.1$ \\
\hline
$r_{3,0}$ &=& $-\frac{22631621}{82944} - \frac{13939}{432} \zeta_3 + 
\frac{4675}{48}\zeta_5 + \frac{107}{24}\pi^2 $ &=& $-166.6 $ \\ 
$r_{3,1}$ &=& $ \left( \frac{107}{8} \right) 
\left(- \frac{208411}{3456} - \frac{20755}{288}\zeta_3 -\frac{355}{48}\zeta_5 
- \frac{137}{192}\pi^2 \right) $ &=& $-2162$ \\
$r_{3,2}$ &=& $ \left( \frac{107}{8} \right)^2 
\left( \frac{694303}{6912} - \frac{119}{4} \zeta_3 + \frac{25}{4}\zeta_5 
- \frac{317}{96}\pi^2 \right) $ 
&=& $6901$ \\
$r_{3,3}$ &=& $ \left( \frac{107}{8} \right)^3 
\left( \frac{985}{108} - \frac{5}{2} \zeta_3 - \frac{11}{24}\pi^2 \right) $ 
&=& $3808$ \\
\hline
\end{tabular}
\end{center}

      In Ref. \cite{h3} it was observed that, at third order in the 
effective-charge (or `FAC') scheme, there is a fixed point with 
${\cal R}^*_{\rm Higgs} \sim a^* \approx 0.15$.   The authors viewed this 
as probably spurious.  Indeed, it is odd that it is only about half the 
size of the frozen couplant found in the $e^+e^-$ case, and so is far 
from the leading-order BZ expectation that $a^* \sim a_0$.  At 4th order 
Ref. \cite{ver} finds that this fixed point is no longer present.  
We have checked that the situation is much the same in optimized perturbation 
theory \cite{opt,kss}.

      The $Q=0$ BZ series in this case is:
\BE
{\cal R}^*_{\rm Higgs} = a_0(1 + 3.05 a_0 + 7.67 a_0^2 + \ldots )
\EE
whose coefficients are considerably larger than in the $e^+e^-$ or 
Bjorken-sum-rule cases (Eqs. (\ref{ree}), (\ref{rbj})).  For a low number 
of flavours the ``corrections'' are as big as the leading term, and 
both the same sign.  (The next coefficient in the expansion is 
$c_{4,-1}-109$ and is still unknown; our estimate (see Eq. (\ref{g3}) 
suggests that it is around $26 \pm 10$.)  One could conclude that this is 
perhaps a case where the BZ expansion breaks down before $n_f=2$, so 
that maybe there is no freezing in this case.  

     We believe, however, the problem is that this quantity is just 
not a useful indicator of massless QCD's infrared behaviour, because of 
the way that quark masses are involved.  For exactly massless quarks 
the hadronic Higgs-decay rate is zero, because the Higgs-quark coupling 
is proportional to quark mass.  (The calculations neglect quark masses 
in the radiative corrections but not, of course, in the overall coupling 
factor.)  If we keep the quark masses finite when we consider, theoretically, 
the limit $M_H \to 0$, we will trivially get zero as soon as the decay becomes 
kinematically forbidden.  To avoid this we would need to consider a limit 
in which $m_q$ tends to zero at least as fast as $M_H$; say 
$m_q \propto (M_H)^{\kappa}$ with $\kappa \ge 1$.  But then 
${\cal R}_{\rm Higgs}$ is not of the form $a(1 + r_1 a + \ldots )$, and 
depends on $\kappa$, making it ill-defined.  These issues do not arise 
for ${\cal R}_{e^+e^-}$ or ${\cal R}_{\rm Bj}$, which are meaningful for 
massless quarks.

\renewcommand{\textfraction}{0.0}
\renewcommand{\floatpagefraction}{.20}
\renewcommand{\topfraction}{1.0}
\renewcommand{\bottomfraction}{1.0}
\renewcommand{\baselinestretch}{1.3}
\setlength{\textheight}{8.5in}
\renewcommand{\thepage}{\arabic{page}}
\setlength{\unitlength}{1in}

\epsfxsize=6.0in
\begin{figure}[t]
 \vskip 1.0 in
 \epsffile[70 159 560 622]{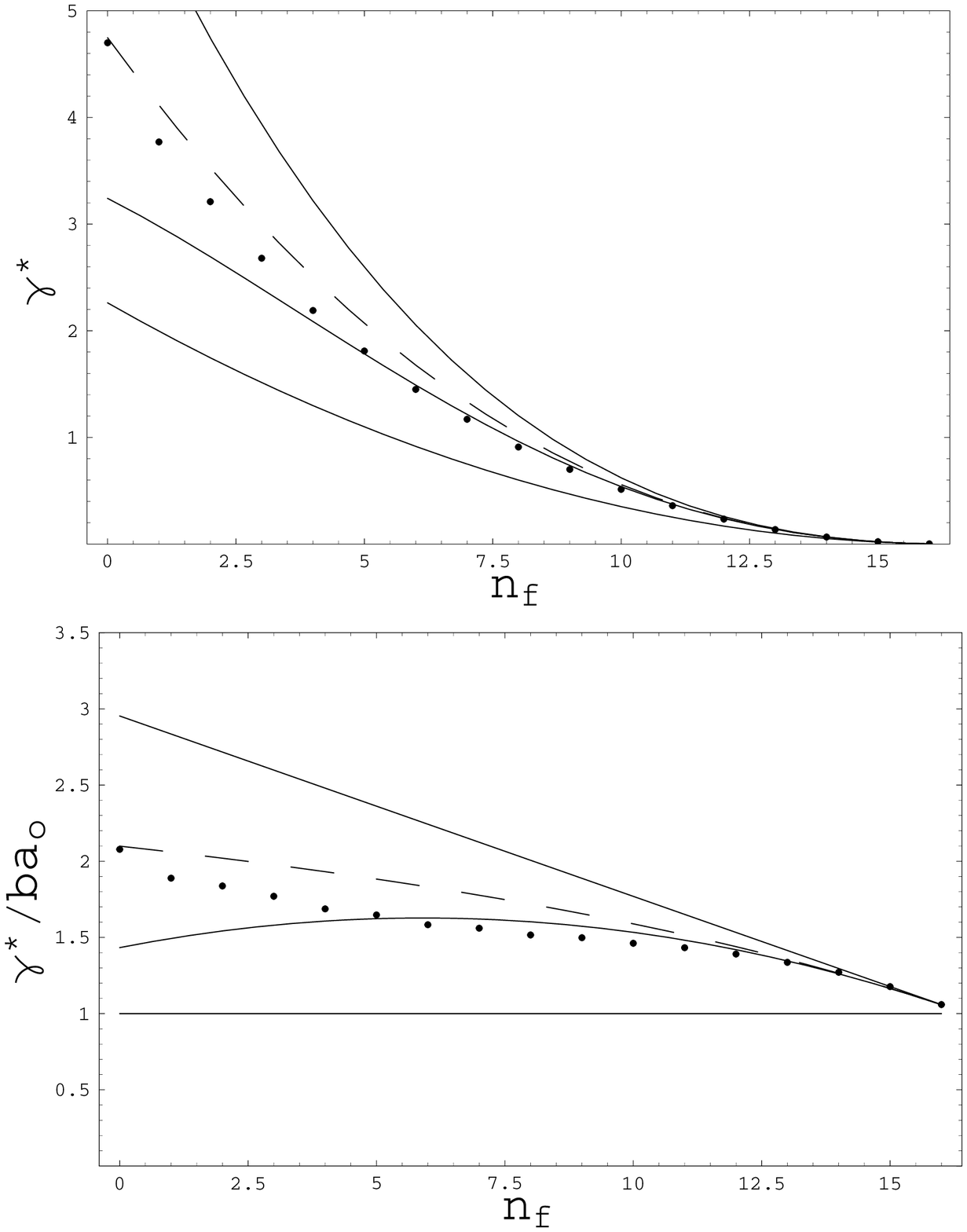}
\end{figure}

\newpage

\mbox{}
\vskip -1in
\begin{picture}(6,.001)(0,1.25)
 \put(1.35,7.0){2nd}
 \put(0.65,6.1){3rd}
 \put(0.55,5.55){1st}

 \put(1.0,3.25){2nd}
 \put(1.0,1.95){3rd}
 \put(1.0,1.45){1st}

\end{picture}
%
\mbox{} 
\vfill
\noindent {\bf Fig. 1.}  
(a) The critical exponent $\gamma^*$ in first, second, and third 
orders of the BZ expansion (lower, upper, and middle solid curves).  
A Pad\'e approximant form of the third-order result is shown as the 
dashed curve.  The dots represent the result of an 
optimized-perturbation-theory analysis [1].  (b)  The same, 
normalized by $1/(b a_0)$; i.e., $\hat{\gamma}^*/a_0$.

\newpage

\epsfxsize=6.0in
\begin{figure}[t]
 \vskip 1.0 in
 \epsffile[70 159 560 622]{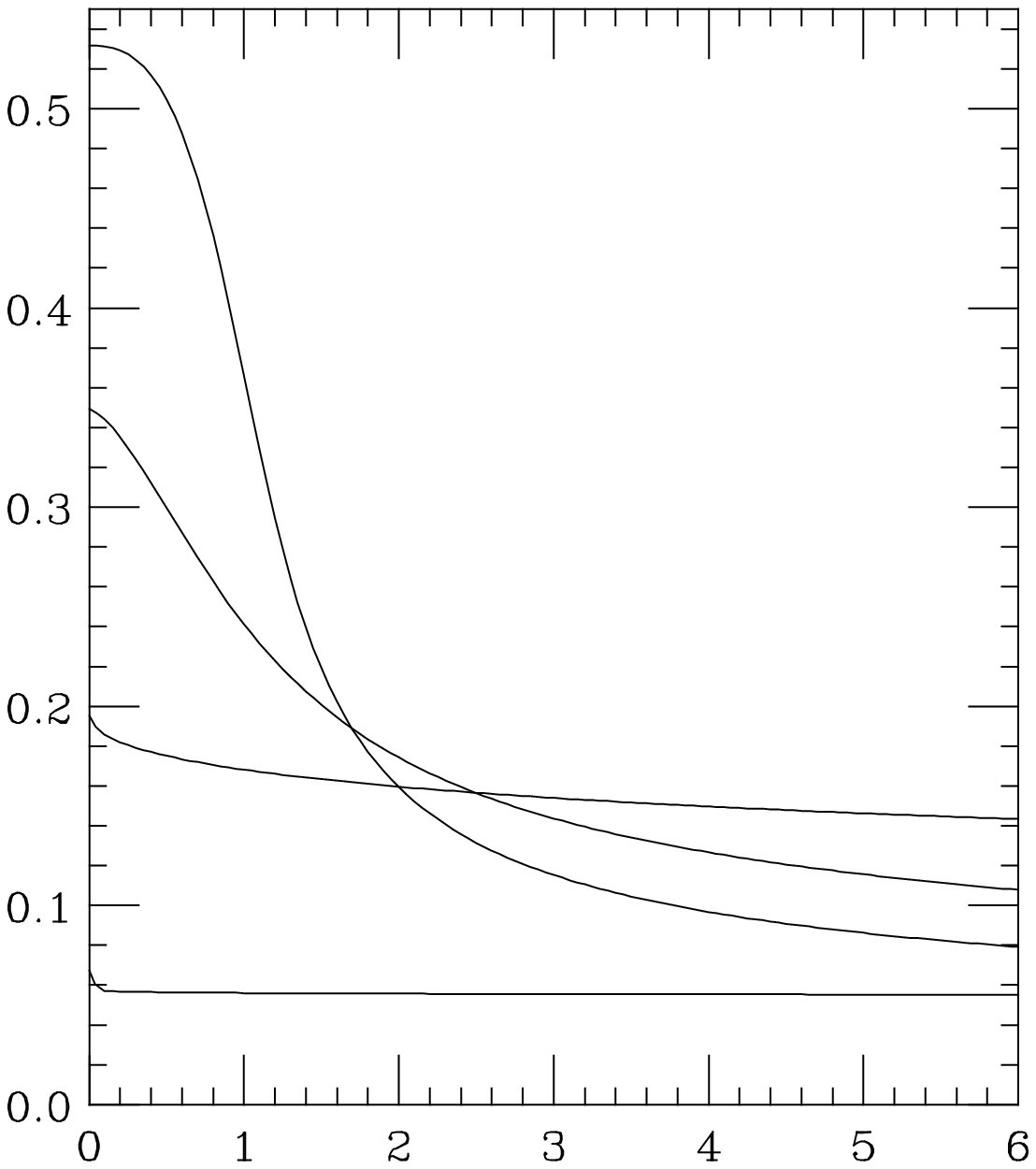}
\end{figure}
\mbox{}
\vskip -1in
\begin{picture}(6,.001)(0,1.25)

 \put(2.8, 2.02){\LARGE $Q/\tilde{\Lambda}_{\rm eff}$}
 \put(0.2,5.10){\LARGE ${\cal R}_{e^+e^-}$}

 \put(1.80,5.35){$2$}
 \put(1.45,4.90){$6$}
 \put(1.40,3.95){$10$}
 \put(1.40,3.03){$14$}

\end{picture}
\mbox{} 
\vfill
\noindent {\bf Fig. 2.}  ${\cal R}_{e^+e^-}$ as a function of 
$Q/\tilde{\Lambda}_{\rm eff}$ to third order in the BZ expansion \\
for $n_f = 14,10,6,2$.
%

\newpage

\epsfxsize=6.0in
\begin{figure}[t]
 \vskip 1.0 in
 \epsffile[70 159 560 622]{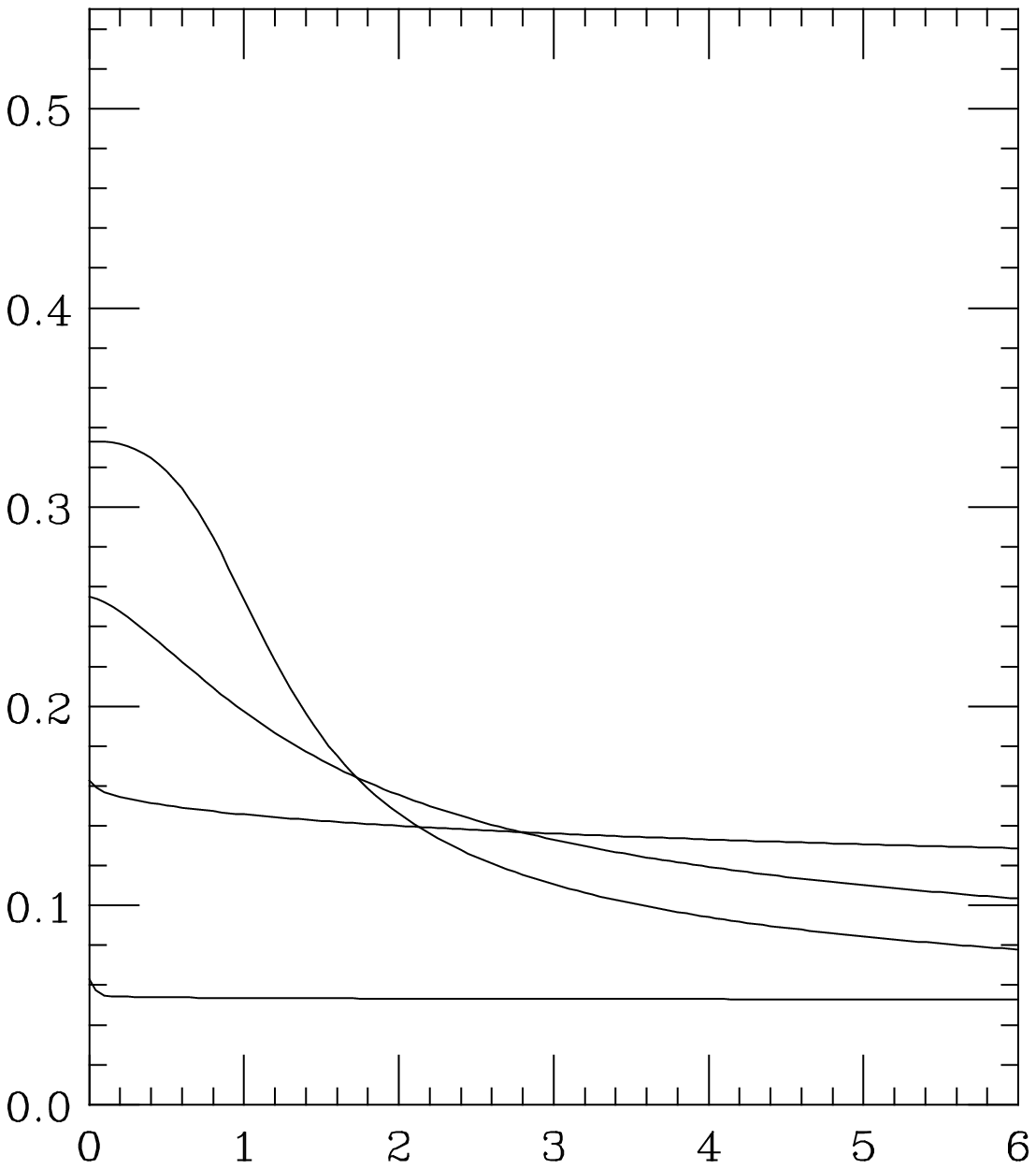}
\end{figure}
%

\mbox{}
\vskip -1in
\begin{picture}(6,.001)(0,1.25)

 \put(2.8, 2.02){\LARGE $Q/\tilde{\Lambda}_{\rm eff}$}
 \put(0.2,5.10){\LARGE ${\cal R}_{\rm Bj}$}

 \put(1.55,4.87){$2$}
 \put(1.46,4.32){$6$}
 \put(1.40,3.77){$10$}
 \put(1.40,3.03){$14$}

\end{picture}
%
\mbox{} 
\vfill
\noindent {\bf Fig. 3.}  ${\cal R}_{\rm Bj}$ as a function of 
$Q/\tilde{\Lambda}_{\rm eff}$ to third order in the BZ expansion \\
for $n_f = 14,10,6,2$.
\end{document}